\documentclass[showkeys]{revtex4}           % referee version: for submission

\usepackage{graphicx,times}             %for PS/EPS graphics inclusion, new

\begin{document}
\title {Spectral lags of flaring events in $LSI +61^{o} $ 303 from RXTE Observations} 
\author{Tamal Sarkar, Samir K Sarkar and Arunava Bhadra\thanks{Email address: aru\_bhadra@yahoo.com}} 
%\email{biplabbijay@rediffmail.com}
%\author{} 
%\author{Arunava Bhadra} 
%\email{aru\_bhadra@yahoo.com}
\affiliation{High Energy \& Cosmic Ray Research Centre, University of North Bengal, Siliguri, WB 734013 INDIA}
%   \title{Spectral lags of flaring events in $LSI +61^{o} $ 303 from RXTE Observations}
%   \subtitle{I. Place Your Subtitle Here}

%   \volnopage{Vol.0 (200x) No.0, 000--000}      %%preserved for Editor. DOn't remove!
%   \setcounter{page}{1}          %%starting page, preserved for Editor. DOn't remove!

%   \author{Tamal Sarkar
%   \inst{1,2}   
% 	\and Samir Sarkar
%	 \inst{1}
%   \and Arunava Bhadra
%	 \inst{1,*}

%   \institute{High Energy \& Cosmic Ray Research Centre, University of North Bengal, Siliguri, WB 734013 INDIA;
%        \and
%             University Science Instrumentation centre, University of North Bengal, Siliguri, WB, 734013, India\\
%        \and
%             Full institute address for the third author\\
%   {*\it aru\_bhadra@yahoo.com}
%	}

%   \date{Received~~2015 October 06; accepted~~2015~~November}

\begin{abstract} This work reports the first discovery of (negative) spectral lags in the X-ray emission below 10 keV from the gamma ray binary $LSI +61^{o} $ 303 during large flaring episodes using the Rossi X-ray Timing Explorer (RXTE) observations. It is found from the RXTE data that during the flares, low energy (3-5 KeV) variations lead the higher energy (8-10 keV) variations by few tens of seconds whereas no significant time lag is observed during the non-flaring states. The observed spectral lag features for flaring events suggest that inverse Compton scattering may be operative, at least in some part of the system. Another possibility is that the sites of particle acceleration may be different for flaring and non-flaring events such as in the microquasar model the flaring radiation may come from hot spots sitting above the black hole while steady state emissions are due to the jets.    
\end{abstract}

%}

%   \authorrunning{Sarkar, Sarkar \& Bhadra }            %author_head in even pages
%   \titlerunning{Spectral lags of flaring events in $LSI +61^{o} $ 303}  % title_head in odd pages

   \maketitle
%% The author head (on even pages) and the title head (on odd pages) will be
%% automatically extracted from \author{} and \title{}. Whenever the title is too long,
%% you will be asked to supply a shorter one by inserting either \authorrunning{} or
%% \titlerunning{} before \maketitle. Anyway, you can specify your own heads.
%%
%%
%% Note: In the following text body of your manuscript, please note several differences from
%%       other major journals:
%% (1) \subsection{Please Capitalize the First Letter of Each Notional Word in Subsection Title}
%% (2) Please Capitalize the First Letter of Each Notional Word in all tables' captions

%
%________________________________________________ sections below
%

\section{Introduction} 
In recent times the gamma ray binary system $LSI +61^{o} $ 303 draws a lot of interest due to the fact that though it is one of the three known gamma-ray binaries detected in the whole energy range from radio to TeV and the system has been observed extensively but the nature of the compact object of the system, particularly whether it is a pulsar (neutron star), or a micro-quasar (black hole) is still not clear. In the former scenario \cite{mar81} the non-thermal emission from the source is expected to be powered by the interaction between the stellar and pulsar winds whereas for the later possibility \cite{tay84} the non-thermal emission is likely to be powered by accretion and jet ejection. 

The presence of an extended jetlike and precessing radio emitting structure \cite{mas01, mas04}  led to microquasar status of the source. However, a rotating elongated (cometary) morphology is noticed in VLBA images \cite{dha06} which is interpreted as due to the interaction between the relativistic wind of a pulsar and wind of companion Be star. Such a scenario receives   support from a recent VLBI observations \cite{mol12}. However, no clear evidence for a jet (a signature for accretion) has been found so far, nor any pulsations have been convincingly detected yet in the deep radio \cite{mcs11} and X-ray searches that could confirm the presence of a pulsar in the system. The emission mechanisms powering the system also remain uncertain. Recently two short bursts have been observed from the direction of $LSI +61^{o} $ 303 with SWIFT-BAT \cite{bar08, tor12} having characteristics typical of those observed in magnetars (strongly magnetized neutron stars) implying that the system may host a magnetar \cite{pap12}. 

Astronomical observations suggest that $LSI +61^{o} $ 303 is a high-mass X-ray binary consisting of a low-mass $[M~(1-4) M_{\odot}]$ compact object orbiting around an early type rapidly rotating B0 Ve main sequence star with a stable equatorial shell along an eccentric $e \sim 0.7$ orbit \cite{gre78}. The radio source GT 0236 +601, is considered to be associated with this Be star, the radio outbursts show a periodicity of about 26.496 days \cite{gre78} and a further modulation of the outburst phase and outburst peak flux with a (super-orbital) period of $~$ 1667 +/- 8 days \cite{gre02}. The same orbital periodicity is also found in other bands - optical \cite{hut81, men89}, infrared \cite{par94}, soft x-rays \cite{par97,tor10} and gamma rays between MeV \cite{mas04, abd09} to TeV \cite{alb06}. TeV gamma rays are, however, detected only close to apastron passage of the compact object in its orbit around the Be star. 

A simple featureless absorbed power law in the X-ray energy band provides an effective fit to the observational spectral data from the source. Such a power law behavior continues in the MeV to GeV energy range with an exponential cutoff at $3.9$ GeV as revealed from the Fermi-LAT observations \cite{had12, abd09}.  An anti-correlation between flux and x-ray photon index was found in a long-term X-ray monitoring campaign of the system by RXTE mission \cite{li11}. Regarding temporal behavior, the X-ray observations conducted on $LSI +61^{o} $ 303 showed extreme variation in the light curve in kilo-second (ks) time scale between different orbital cycles: the flux is found to vary by a factor of about 3 over a single orbital cycle with a peak flux between orbital phases 0.4 to 0.6 \cite{gol95, tay96}. A sharp decrease in the flux of the order of a factor of 3 over a time period of a few thousand seconds at an orbital phase 0.61 was noticed from an observational campaign of the system using XMM-Newton \cite{sid06}. It was further found that the hardness ratio changes with the change of the flux. The presence of kilo-second scale mini flares in the source, with emission increasing by a factor of 2 over a timespan of roughly 1 hour along with a correlation between harder emission and increased flux were divulged from the Chandra observations (0.3 -10keV) of $LSI +61^{o} $ 303 near orbital phase 0.04 \cite{par07}. 

A long time monitoring of the system was performed by the RXTE mission covering the period between 2006 October and 2010 September that includes $42$ adjacent cycles of the system orbital motion \cite{li11,li12}. Besides various important spectral and temporal behavior of the system the RXTE observations led to the discovery of sub-kilosecond scale flaring episodes from the system. Total five large flaring episodes were revealed from the RXTE observations which showed that the X-ray emission from the source changing by up to a factor of 6 over timescales of few tens of seconds and the doubling times is as fast as 2 seconds \cite{smi09, li11}. 

In this article, exploiting the RXTE observations of $LSI +61^{o} $ 303 we have presented the results of the spectral time lag i.e. the difference in time of arrival between high- and low-energy photons in the large flaring episodes. We shall show the evidence that in the flaring episodes the hard x-ray peak flare emission lags behind the soft x-ray emission with a time delay of few tens of seconds. In contrast during the steady (non-flaring) states, spectral lag of the system is found consistent with zero lag. The implications of the present findings on the models of $LSI +61^{o} $ 303 are discussed.  
% \end{document}
\section{Data reduction}

The RXTE Proportional Counter Array made a long-term monitoring of $LSI +61^{o} $ 303 covering 42 adjacent cycles of the  26.496 day binary orbital period between 2006 October and 2010 September and detected five flaring episodes of few hundred seconds scale. To study the spectral lag behavior of all the flares observed with RXTE and for spectral analysis, we follow the same data analysis method adopted in \cite{smi09, li12} i.e., for spectral lag study light curves are generated using the \lq\lq Good Xenon\rq\rq mode (resolution less than 16 s) from RXTE/PCA data whereas for spectral analysis the \lq Standard 2 mode was exploited. Data reduction was performed using HEASoft 6.12. We filtered the data using the standard RXTE/PCA criteria. The pcabackest has been used to generate simulated background spectra/light curve (using standard 2 mode) and PCA response matrices are obtained with pcarsp. The background lightcurve was subtracted from the Good Xenon observations by utilizing the FTOOLS routine lcmath. For the study of spectral lag in this work we derived the light curves for two widely different energy intervals, 3-5 keV and 8-10 keV.  

A proper choice of time binning is very important for extraction of spectral lags. Presumably, by changing the time binning of the light curve one is affecting the signal-to-noise ratio. For very small time binning the signal-to-noise ratio may become very small. On the other hand the sought-after information from the light curve may be lost due to use of overly large time bin sizes. Considering that the five flares observed by RXTE are binned in 5 s intervals. To estimate spectral lags in non-flaring (steady) episodes we select data sets for all the observations from event ID 54345 to 54705 (total 141 observations) covering all the flaring episodes. Applying the cross-correlation function (CCF) technique, as discussed below, we estimated lags between flare light curve at two different energies ($3-5$ and $8-10$ keV) for each observation. 

For the study of spectral behavior again we follow the procedure adopted in \cite{smi09} i.e. the standard quality criteria cuts are applied while selecting data, the standard 2 mode (129 channels) is considered, the background spectra are simulated with pcabackest, the PCA response matrices are created using pcarsp and finally XSPEC12 is exploited to fit the background subtracted spectra in the energy interval 3-10 keV.    

\section{Results}
Five large flaring episodes, each of few hundred seconds have been observed by the RXTE mission (table 1). The MJD 54356 is reported as the most powerful of the observed flares with flux $7.2 \pm 0.2 \times 10^{-11}$ $erg\; s^{-1}\; cm^{-2}$ which is larger than 3 times of the average flux in the system. In contrast the flux level of the system for non-flaring events is found to modulate typically between $0.5 \times 10^{-11}$ to $2.0 \times 10^{-11}$ $erg\; s^{-1}\; cm^{-2}$ \cite{smi09}. Except one (MJD 54372) all the flares are observed between $0.6$ to $0.9$ orbital phase bins of LS I +61◦303 setting the phase zero at JD 2,443,366.775 following the orbital solution by \cite{cas05}. During flaring episodes $4$ and $5$ the spectral index varies with time and the flux exhibits a linear anti-correlation with spectral index as reported by \cite{li11}. 

An important question to address before spectral lag analysis is whether any real variability exist in the light-curves considered here i.e. the observed flares are not caused by the same stochastic process that produces variability in the rest of the lightcurve. To examine this we first compute the count rate in each observations (the RXTE mission took nearly one kilosecond exposures every day over a period of four years) and plot the time variation of the count rate in two different energy bands, $3-5$ keV and $8-10$ keV as displayed in figure 1 over the period MJD 54340 to MJD 54390. The count rates in the event IDs MJD 54356, 54358 and 54372 are found more than 3$\sigma$ higher than the average count rate and thereby the events are identified as flaring episodes. We have also performed the Kolmogorov-Smirnov (KS) test with Lilliefors correction \cite{lil67} to check the variability. For this also we consider the count rate for each observations over the period MJD 54340 to MJD 54390. The result is shown in figure 2. Assuming the null distribution is to be normal, the null hypothesis (the signals are not flare but simply Gaussian noise) is rejected with p value less than 0.01 and 0.05 respectively in energy bands, $3-5$ keV and $8-10$ keV.

\begin{figure} 
%\plottwo{54356a.ps}{54356a_color.ps}
\includegraphics[width=84mm]{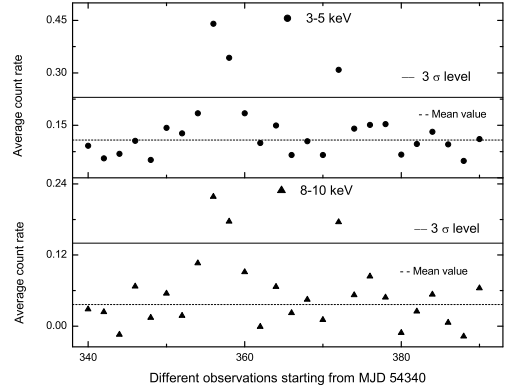}
\caption{ Time (in modified Julian day) variation of the count rate over the period MJD 54340 to MJD 54390 for two x-ray energies 3-5 keV (upper panel) and 8-10 keV (lower panel). The dotted line denotes the average count rate over the stated time period and the solid line represents the 3$\sigma$ count rate above the mean. The three flaring episodes, MJD 54356, MJD 54358 and MJD 54372 were detected with more than 3$\sigma$ significance level. }
\label{fig1}
\end{figure}

\begin{figure} 
%\plottwo{54356a.ps}{54356a_color.ps}
\includegraphics[width=84mm]{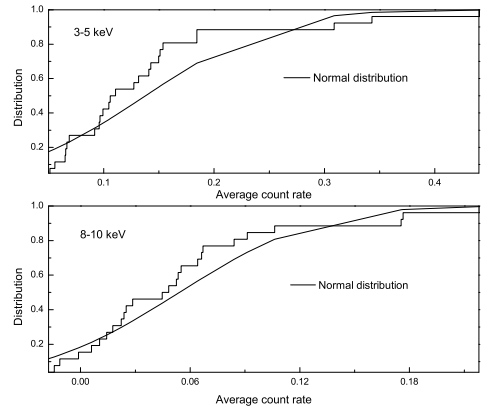}
\caption{The KS test with Lilliefors correction for the count rate over the period MJD 54340 to MJD 54390 for two x-ray energies 3-5 keV (upper panel) and 8-10 keV (lower panel). The p values are found $<0.01$ and $<0.05$ respectively for the energy bands 3-5 keV and 8-10 keV. }
\label{fig2}
\end{figure}

Each flaring episodes contains several sub-flares with variability in the time scale of few ten seconds as has been (originally) demonstrated by Smith et al \cite{smi09} and Li et al \cite{li11} from RXTE data. However, not all the sub-flares are well structured having clear rise and fall \cite{smi09}. To examine that the sub-flares within flaring episodes are real, not noise we again employ the Kolmogorov-Smirnov (KS) test with Lilliefors correction \cite{lil67} as shown in figure 3 for the flaring event id MJD 54372. The confidence level of the variability found from KS test with Lilliefors correction for MJD 54356, MJD 54358 and MJD 54372 are $>93 \%$, $>95 \%$ and $>99 \%$ respectively in the energy band $3-5$ keV and $>99\%$, $>92\%$ and $>99\%$ respectively in the energy band $8-10$ keV. The KS test of the flaring IDs thereby indicates that the sub-flares within flaring episodes are also real effects. We, therefore, proceed to study spectral lag for the flares.  

\begin{figure} 
%\plottwo{54356a.ps}{54356a_color.ps}
\includegraphics[width=84mm]{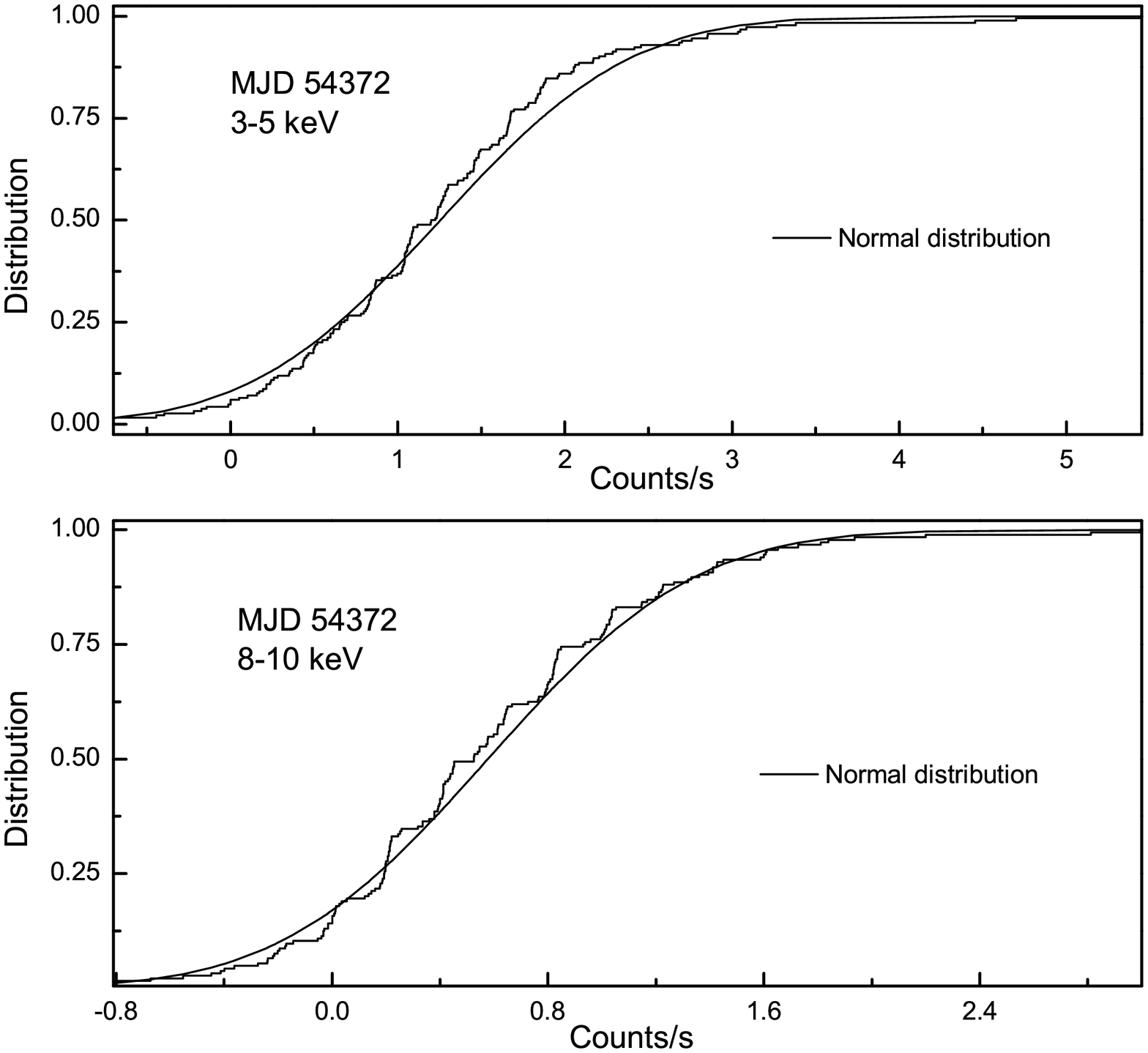}
\caption{The KS test with Lilliefors correction for the flaring event MJD 54372. The upper and lower panels correspond to energy window $3-5$ keV and $8-10$ keV respectively. }
\label{fig3}
\end{figure}

\begin{figure} 
%\plottwo{54356a.ps}{54356a_color.ps}
\includegraphics[width=84mm]{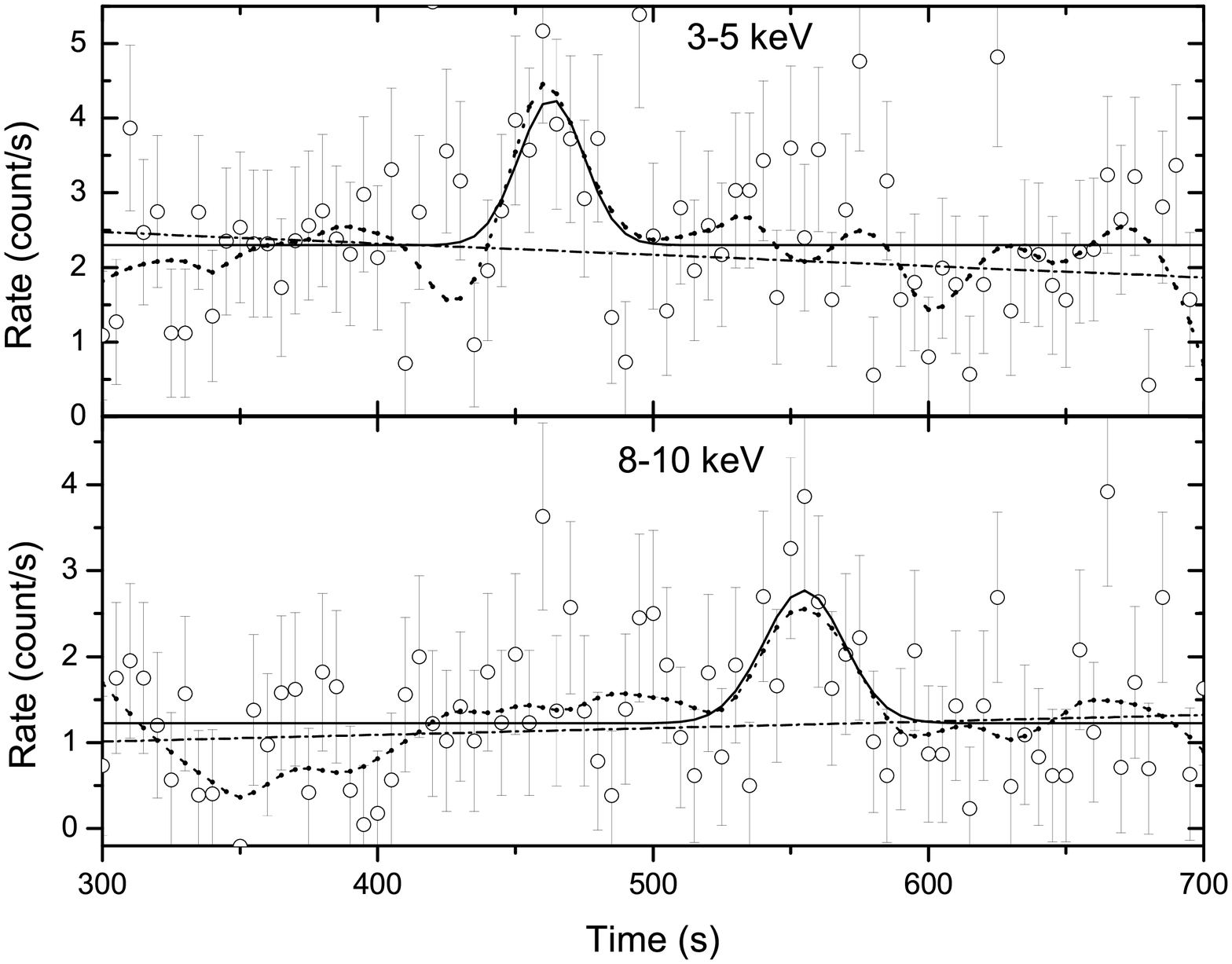}
\caption{The light curve of the flaring event 54356 over the time window 300 -700 sec of the observation. The solid and dotted lines represent respectively the Gaussian fitting of the sub-flare and the smoothed line after application of the Loess method with the parameter $\alpha=0.2$. }
\label{fig4}
\end{figure}

For the purpose of estimating spectral lag we need the presence of at least one well defined flare having clear rise and fall in light-curves of both the considered energy bands ($3-5$ keV and $8-10$ keV). After imposing such conditions we found only few flaring events such as those given in Figures 4 - 5 for the event ID MJD 54356 and MJD 54372 respectively.

\begin{figure}
\includegraphics[width=84mm]{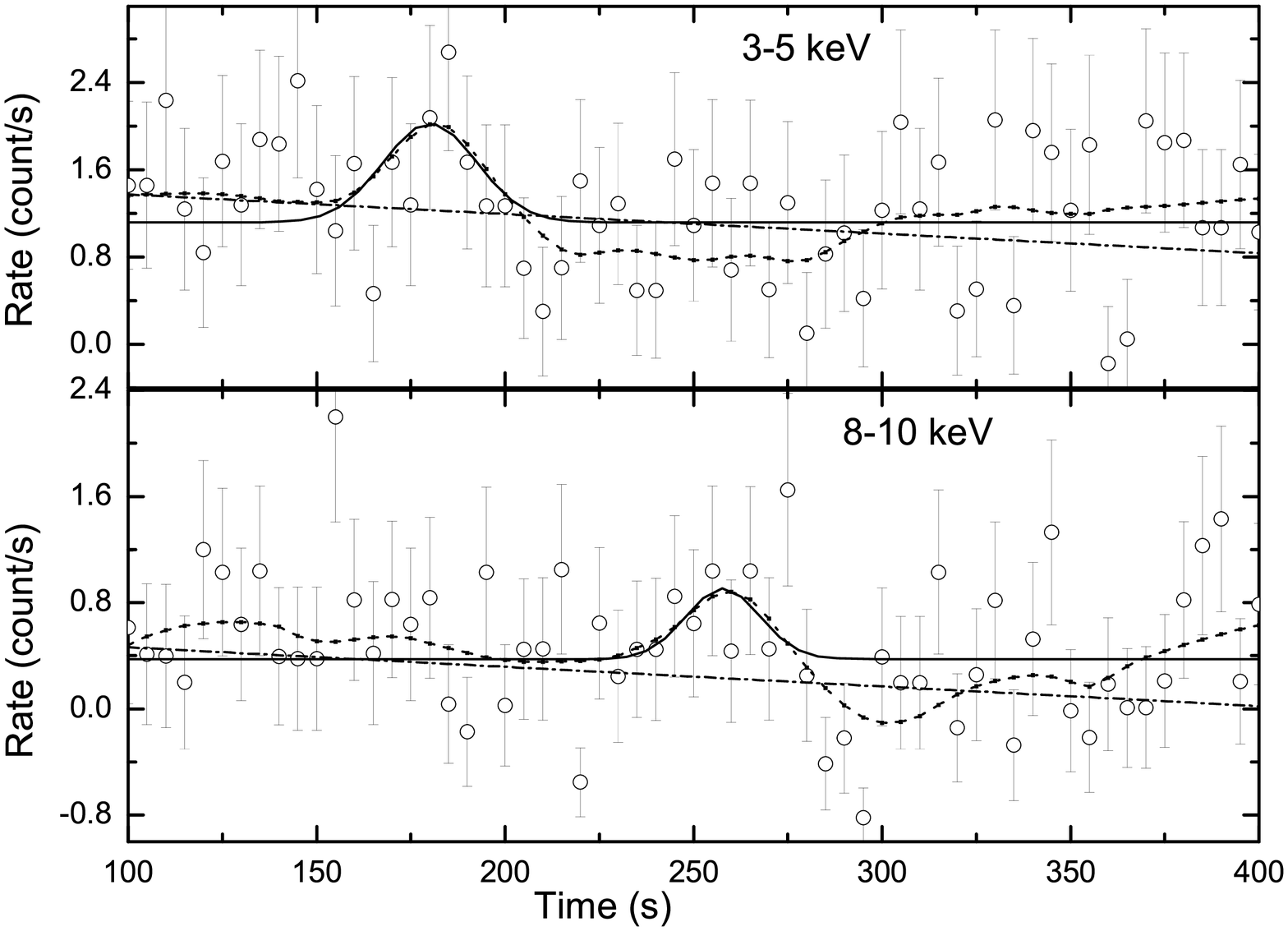}
\caption{Same as Figure 4 but for the flaring event MJD 54372 over the time window 100-400 sec of the observation.\label{fig5}}
\end{figure}

Generally spectral lag is extracted either from pulse peak - fit method \cite{nor05, hak08}, or through cross-correlation function (CCF) analysis method \cite{lin93, fen95, che95, nor00}. The former approach requires a certain pulse model for the pulses in the light curve. In the CCF analysis approach, the time delay corresponding to the global maximum of the CCF gives the spectral lag. The CCF between two energy bands have been calculated using the crosscor tool in Xrones (HEASOFT) and set the normalization parameter (option 0) such that the count rates are normalized to the average rate. 

\begin{table}
%\begin{minipage}
\begin{center}
\caption{\it The details of the Flaring events observed by RXTE mission over the energy range 3-10 keV. The phases have been determined setting the phase zero at JD 2,443,366.775. }
\begin{tabular}{crrr}
\hline
Sl no. & Event ID & phase &  flux  \\
   &     MJD      &        &  (erg $s^{-1} cm^{-2}$)  \\
\hline
1 &  54356 & 0.787 & $7.2\pm 0.2 \times 10^{-11}$\\
2 & 54358 & 0.861 & $3.5^{+0.1}_{-0.2} \times 10^{-11}$ \\
3 & 54372 & 0.379 & $4.9^{+0.1}_{-0.2} \times 10^{-11}$ \\
4 & 54670.84444 & 0.651 & $(1-5) \times 10^{-11}$\\
5 & 54699.653333 & 0.739 & $(2.5 -10) \times 10^{-11}$ \\

\hline
\end{tabular}
\end{center}
%\end{minipage}
\end{table}

We adopt both the mentioned techniques for estimation of spectral lag of the flaring events. 

We first apply the direct CCF method. When we considered a longer time interval such as $300 - 700$ seconds for event ID MJD $54356$,  the significance of CCF is small, exhibiting noise like behavior which cannot be satisfactorily fitted by Gaussian or other regular function. This is probably because here the light curve over a longer time interval is characterized by many \lq emission events \rq rising or decaying on different time scales thereby diluting the spectral lag characteristics. When we select marginal time intervals just encircling a flare in the stated two energy bands (i.e. the time interval contains the flare in both the energy bands), we got more regular CCF. For instance, we choose the time interval $420 - 600$ s in the case of MJD 54356 which contains relatively higher mean variance in both the energy range. After selection of time intervals spectral lags are obtained as the time delay corresponding to the global maximum of cross correlation function. Gaussian fitting is used to locate the global maximum as shown in figures 6-10 for the five flares. The estimated spectral lags are displayed in Table 2. Note that even after such time interval selection, we have not got a regular shape CCF for the flare in the event ID MJD 54358, neither we got reasonable significance of CCF.  When we took smaller time bins, no regular behavior of CCF has been observed. The reason is that the spectral delay is so much that the (small) time window does not contain the flare in both soft and hard band light curves. So we come to the conclusion that in the present case spectral lag occurs only for well defined flares, not for the continuum. The results are, however, robust in the sense that small change of the time window (or sliding of the time window) by few tens of seconds does not affect the spectral lag results much. For instance in the case of MJD 54356, the choice of the time window $345 - 550$ seconds give spectral lag $16.8 \pm 1.02$ seconds wheras the time window $450 - 700$ produces spectral lag  $16.18 \pm 3.18$ seconds which are almost the same to what we get $-16.2 \pm 3.8$ for the original time window $420-600$ seconds. However, the significance of CCF are smaller for such other choices of the time window.   

\begin{figure}
\includegraphics[width=84mm]{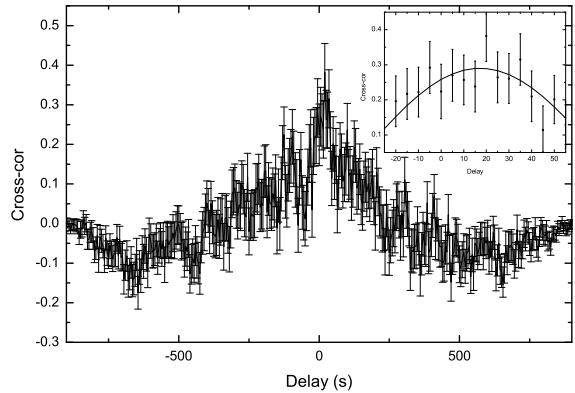}
\caption{The CCF between $8-10$ keV and $3-5$ keV light curves as a function of time delay for a flare in event MJD 54356 and the Gaussian fit of the peak region of the CCF.\label{fig6}}
\end{figure}

\begin{figure}
\includegraphics[width=84mm]{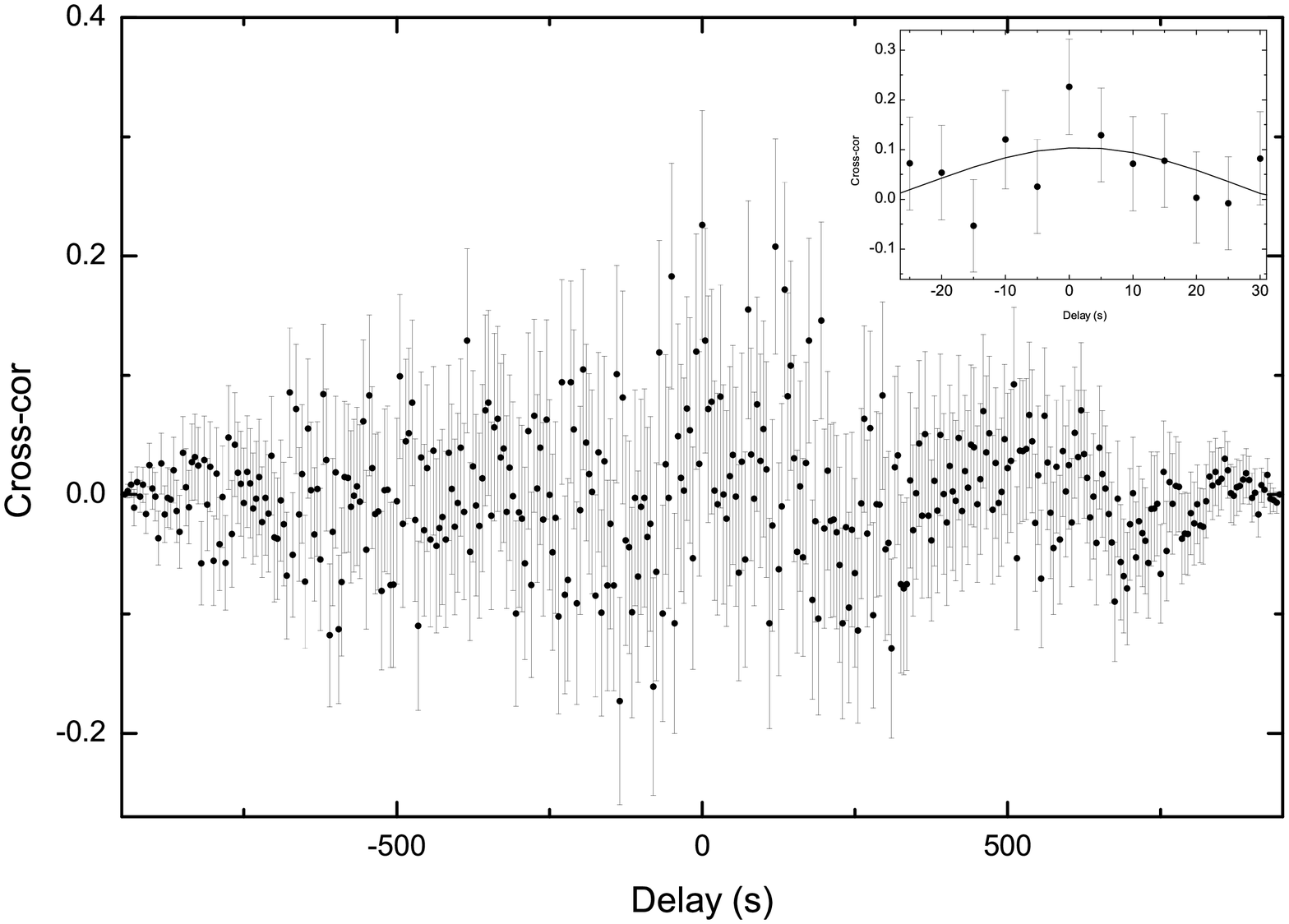}
\caption{Same as Figure 6 but for the flaring event MJD 54358.\label{fig7}}
\end{figure}

\begin{figure}
\includegraphics[width=84mm]{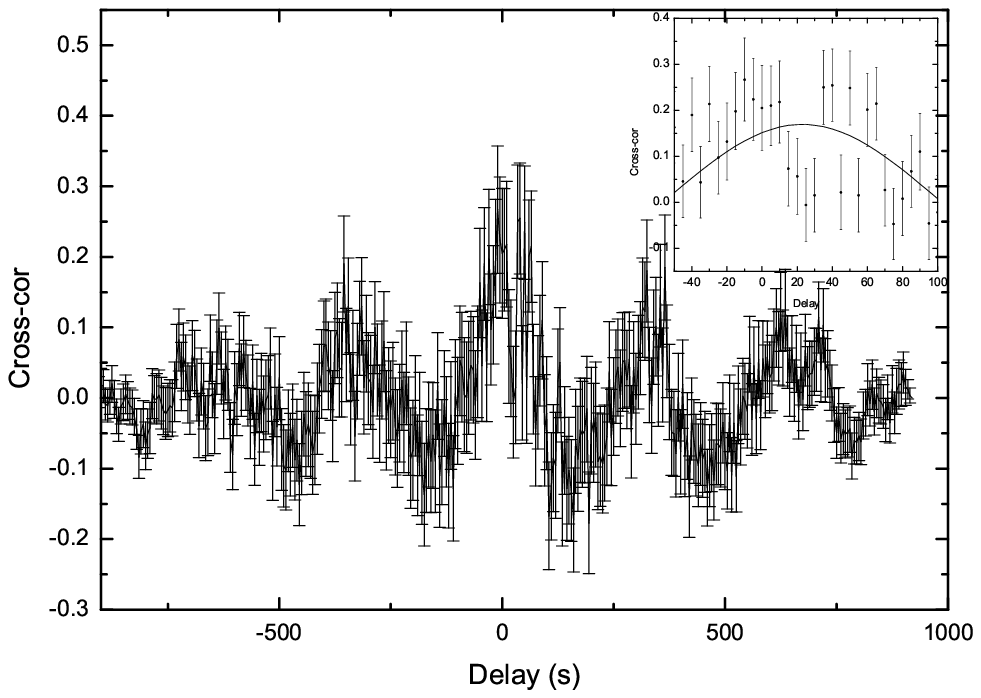}
\caption{Same as Figure 6 but for the flaring event MJD 54372.\label{fig8}}
\end{figure}

\begin{figure}
\includegraphics[width=84mm]{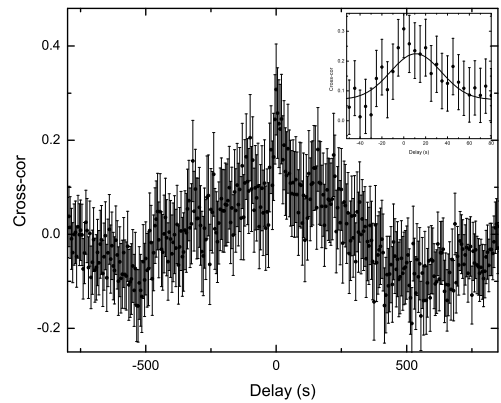}
\caption{Same as Figure 6 but for the flaring event MJD 54670.844444.\label{fig9}}
\end{figure}

\begin{figure}
\includegraphics[width=84mm]{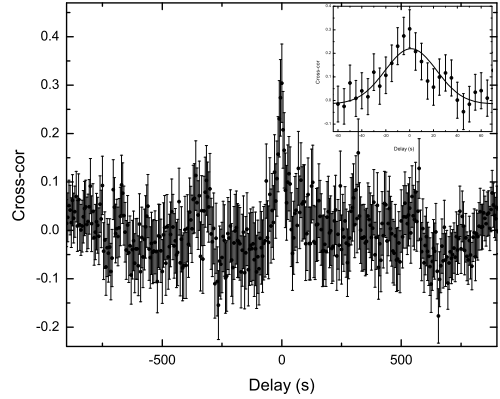}	
\caption{Same as Figure 6 but for the flaring event MJD 54699.653333.\label{fig10}}
\end{figure}

\begin{table}
\begin{center}
\caption{\it Spectral lags between light curves of $8-10$ keV and $3-5$ keV for the flaring events}
\begin{tabular}{crr}
%\tabletypesize{\scriptsize}
\hline
Flaring events & Event ID     & Spectral lag       \\
               &              & by CCF             \\
               &   (MJD)      & (in s)             \\
\hline
1              &54356         &-16.2 $\pm$ 3.8     \\
2              &54372         &-23.0 $\pm$ 6.6    \\
3              &54358         &-2.2 $\pm$ 1.8    \\
4              &54670.844444  &-5.8 $\pm$ 0.7     \\
5              &54699.653333  &-1.1 $\pm$ 0.3      \\

\hline
\end{tabular}
\end{center}
\end{table}

To estimate the spectral lags precisely for the well defined flares, we first smoothed the light curves of different energy bands using the \lq Loess method\rq \cite{cle79,cle88} and then apply the CCF technique. It has been found that the use of Loess filter provides a more reliable estimation of spectral lags for GRBs \cite{liz12}. Basically it is a bi-variate smoother procedure for drawing a smooth curve through a scatter diagram. In this method, the smooth curve is drawn in such a way that the value of smoothed function at a particular location along the x-axis is determined only by the points in that vicinity and locally, the curve minimize the variance of the residuals. The Loess filter is characterized by a smoothing factor $\alpha$ that determines the smooth-span (in a data set of total N points, each smoothed value at the center of the smoothing span will be generated by a 2nd degree polynomial regression using linear least squares fitting with $N\alpha$ data points). Thus the light curve will be less smoothed if $\alpha$ is small and vice versa. By trial \& error we take $\alpha=0.2$ which gives best fitting to the data points. We observed that for $\alpha \ge 0.3$ the regression function start producing periodicity which is not present in the data. The flares or rather flaring pulses are clearly picked up by the Loess technique as can be seen, for examples, in figures 4 and 5 for the flaring events 54356 and 54372 respectively. Then we calculate the CCF between the smoothed curves at two energies as a function of time delay. For instance the CCF between $8-10$ keV and $3-5$ keV for the flares of event ID 54356 and event ID 54372 as displayed in figures 4 and 5 are shown in figures 11 and 12. Subsequently we estimated the spectral lag by Gaussian fitting of the region around the global maximum of the CCF. The calculated spectral lags by this technique are also displayed in table 3. For $\alpha =0.25$ we get almost the same spectral lag to that obtain with $\alpha =0.2$.

\begin{figure}
\includegraphics[width=84mm]{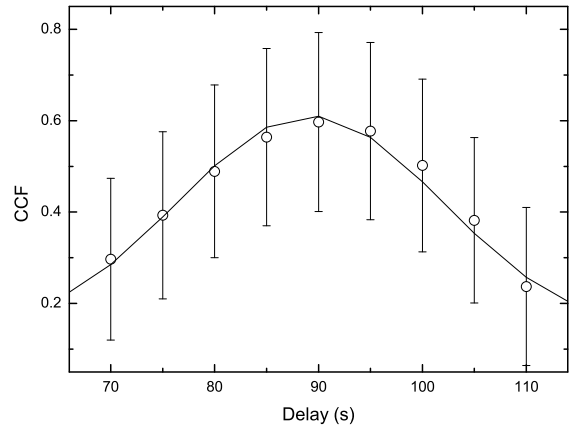}
\caption{Same as Figure 6 but after Loess smoothing.\label{fig11}}
\end{figure}

\begin{figure}
\includegraphics[width=84mm]{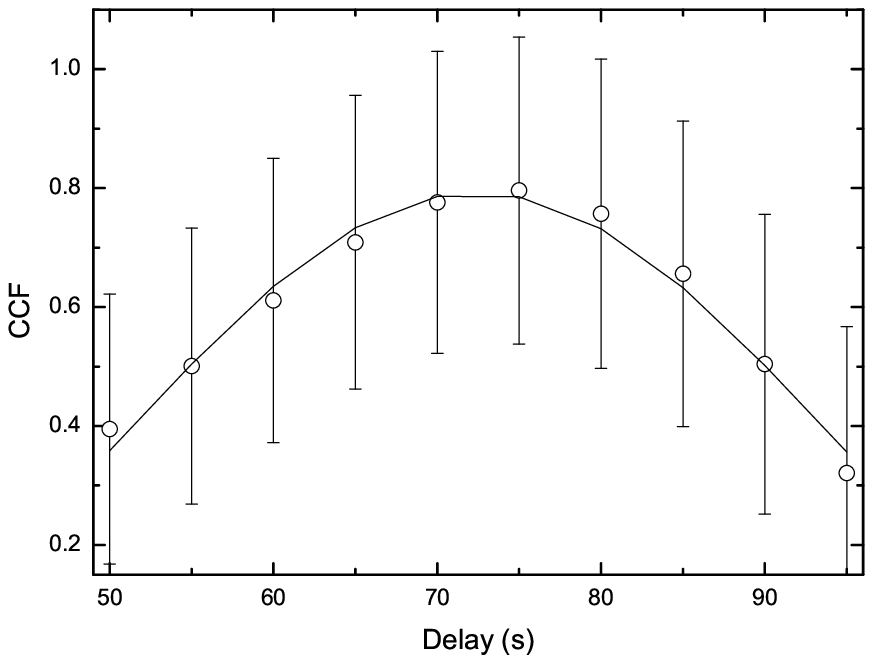}
\caption{Same as Figure 6 but for the flaring event MJD 54372.\label{fig12}}
\end{figure}

\begin{table}
\begin{center}
\caption{\it Spectral lags between light curves of $8-10$ keV and $3-5$ keV for flaring events}
\begin{tabular}{crrr}
%\tabletypesize{\scriptsize}
\hline
Flaring        & Event ID     & Spectral lag by      &  Spectral lag  \\
events         &              & CCF after Loess & from Gaussian fitting   \\
               &   (MJD)      & smoothing  (in s)             \\
\hline
1              &54356         &-89.2 $\pm$ 0.8   & - 91.7 $\pm$ 2.7   \\
2              &54358         &-101.0 $\pm$ 0.6  & -110.1 $\pm$ 4.6   \\
3              &54372         &-72.5 $\pm$ 0.4   & -79.8 $\pm$ 4.7   \\
4              &54670.844444  &-12.0 $\pm$ 0.6   & - 19.1 $\pm$ 4.3   \\
5              &54699.653333  &-1.3 $\pm$ 0.2    & - 1.2 $\pm$ 2.9  \\

\hline
\end{tabular}
\end{center}
\end{table}

In the absence of any definite pulse model for flaring events, we fit each of the flares at different energy bands with Gaussian pulse. For instances, the fitted pulses (the solid lines) are displayed in the figures 4 and 5 in the energy bands $8-10$ keV and $3-5$ keV for the flares in observation ID 54356 and 54372. Here to ensure variability in light curves as well as to examine whether a Gaussian function reasonably describes the flare data we fit the flares data with both straight line and Gaussian curve and subsequently both chi-square and F-test have been carried out. The figure 13 demonstrates such fitting for the flares in event ID MJD 54356. For Gaussian fitting the $\chi^{2}$ values are found $0.88$ and $0.61$ (with 5 second binning) respectively for $3-5$ keV and $8-10$keV respectively and for 2 second binning, they become $0.78$ and $1.10$ respectively for energy bands $3-5$ keV and $8-10$ keV which imply that Gaussian distribution describe the data well. For MJD 54356 the F values are found $1.65$ and $1.28$ respectively for energy bands $3-5$ keV and $8-10$ keV when Gaussian fitted data and observed data are considered as samples which are much smaller than the corresponding critical values of F ($2.22$ and $2.27$) unlike the case of straight line fitting. For energy bands $3-5$ keV and $8-10$ keV the F values are $114.0$ and $9.08$ respectively when Gaussian fitted data and straight line fitted data of MJD 54372 are considered as samples. The corresponding critical values of F are $2.27$ and $2.27$). Thus we see that the observed flares can be reasonably describe by Gaussian function. From the fitted parameters we straightway evaluate the spectral lag between $8-10$ keV and $3-5$ keV energy bands and are also given in table 3. As we mentioned in the section 2, the choice of bin size is very important for studying the flares and thereby we have very limited scope to change the bin size. We have checked that for two bin sizes, 2 and 5 seconds, the Gaussin description of the sub-flares. It is found that the estimated parameters from Gaussian fitting, particularly the mean and standard deviation in two different bin sizes are quite close. For the spectral lag the important parameter is the mean (xc). For 2 second bin, xc equals to $463.7 \pm 1.5$ and $559.7 \pm 7.75$ respectively for $3-5$ keV and $8-10$ keV whereas for 5 seconds bin xc becomes $461.9 \pm 1.1$ and $555.5 \pm 11.6$ for $3-5$ keV and $8-10$ keV respectively.  So the estimated spectral lag is not affected much by the stated change in bin size in the present case.

\begin{figure}
\includegraphics[width=84mm]{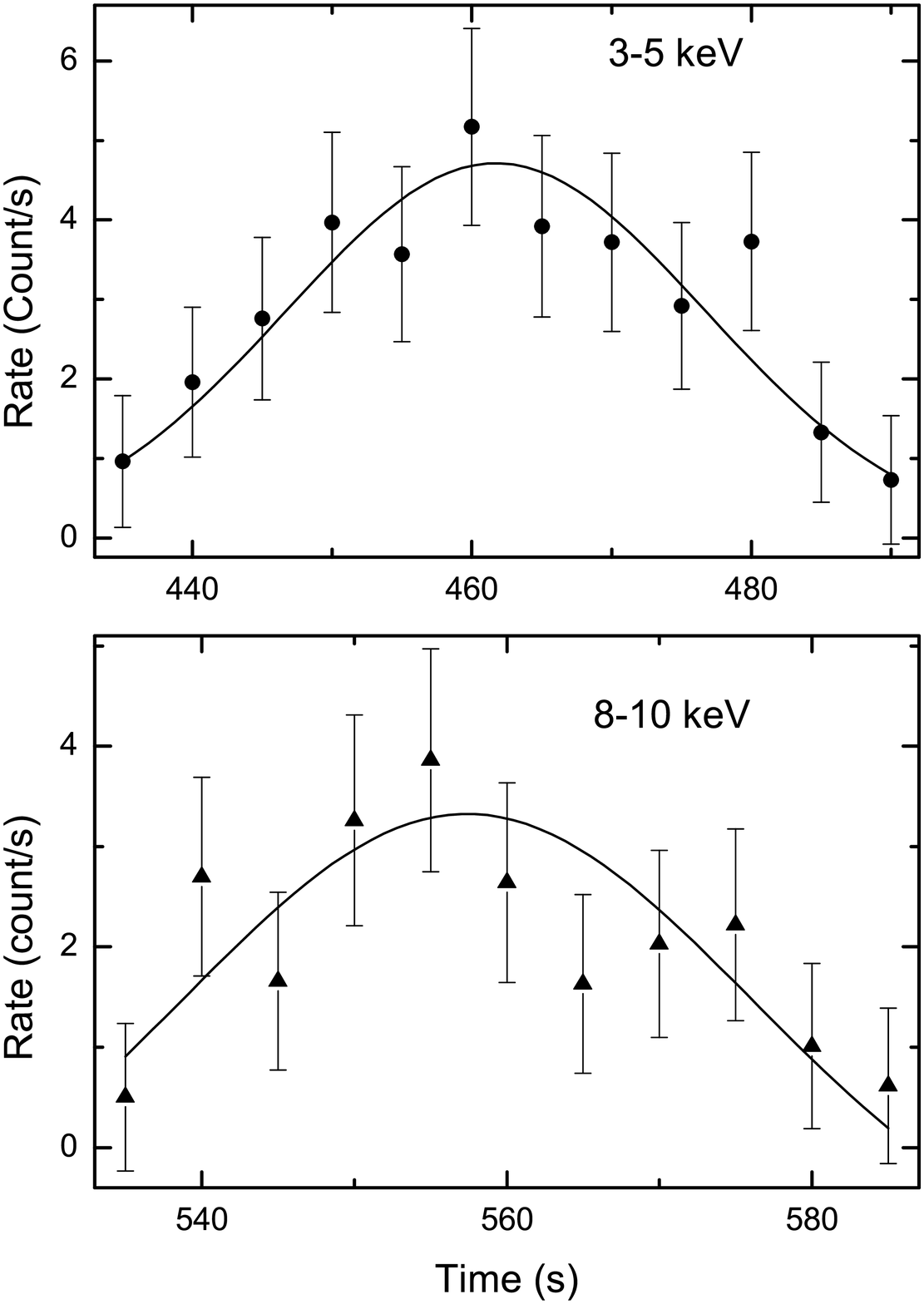}
\caption{Gaussian fitting of flare data in event MJD 54356 for two energy range $3-5$ keV (upper panel) and $8-10$ keV (lower panel). \label{fig13}}
\end{figure}

The lag values obtained by the different techniques (CCF after Loess smoothing and pulse fitting) are found consistent to each other in all the cases. The straightforward CCF gives relatively smaller value of lags. The difference in the lags in CCF analysis before and after the smoothing seems due to noise. A well-defined flare is one which has clear rise and fall. In the case of noise one point may go up, the next point may go down. When LOESS method of smoothing is applied, at each point in the data set a low-degree polynomial is fitted using weighted least squares, giving more weight to points near the point and less weight to points further away. The value of the regression function for the point is then obtained by evaluating the local polynomial. So fluctuations on both sides across the mean constant flux line are somewhat nullified with the use of LOESS smoothing whereas the true flares are singled out. The CCF approach requires pulses in different energy bands to compare with and finding out the lag, if any. From constant flux lines one cannot determine lag. So CCF after the LOESS smoothing gives lag in flares of two different energy bands whereas direct CCF considers the variability due to noise also while evaluating lag. That is why direct CCF in the present case underestimates the lag. 
But importantly all the methods suggest spectral lag. We noted that the spectral lags for the first three flaring episodes are substantially higher than the last two.

Spectral lag is usually caused by spectral hardening between the two spectra though such a logic is not valid always \cite{roy14}. The study of \lq hardness ratio \rq (HR) provides a model-independent method to study the spectral hardening of a source. We calculate hardness ratios for the flares between light curve corresponding to $3-5$ keV and $8-10$ keV. The variation of hardness ration with time is displayed in figure 14 for MJD 54356. However, due to large error bars associated with the points and limited statistics, no firm conclusion could be drawn from the time variation of hardness ratio on spectral hardening. 
 
\begin{figure}
\includegraphics[width=84mm]{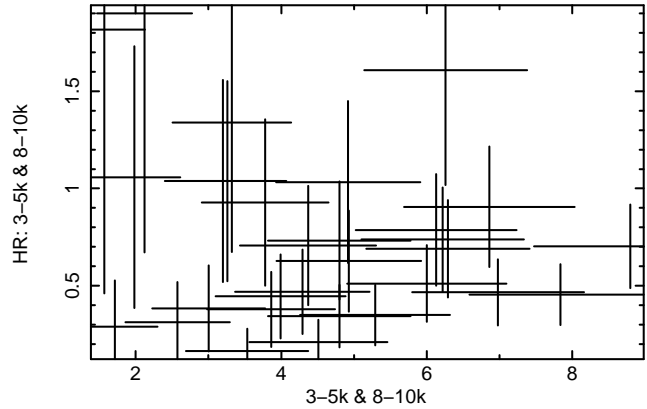}
\caption{Variation of hardness ratio with time for the event MJD54356. \label{fig14}}
\end{figure}

Spectral lag estimation in the steady emission state is much difficult as there is no well defined spectral variation available for comparison between two different energy bands. So we considered the light curves during the same time duration of 200 s at two different energies ($8-10$ and $3-5$ keV) for each observation ID and applied the direct CCF technique. Finally the spectral lags are extracted from Gaussian fit to the CCF (plotted as a function of time delay). However, we found that significance of the CCF for non-flaring events are quite low, neither they have any regular shape as may be seen from figures 15 and 16 for the event id MJD 54352 and 54376. If we still fit the CCFs by Gaussian function, we found that the average spectral lag for the 141 observations between the event ID 54345 and 54705 is $0.01 \pm 0.04$ s and the maximum spectral lags from direct CCF  out of the 141 observations in the steady state are $4.9 \pm 1.9$ and $4.9 \pm 1.5$ s corresponding to the event ID $54352$ and $54385$ respectively. Here it is worthful to say that due to poor correlation no reliable estimate of lag is possible for non-flaring events from CCF studies. If Loess smoothing is applied better correlation between lightcurves of two energies is noticed for non-flaring events also as shown in figure 17 for the event ID MJD 54352 for instance (the corresponding lag becomes $0.4 \pm 0.3$ s). On application of Loess method it is found that spectral lags of all the steady emission events are more or less consistent with zero lag value.  
\begin{figure}
\includegraphics[width=84mm]{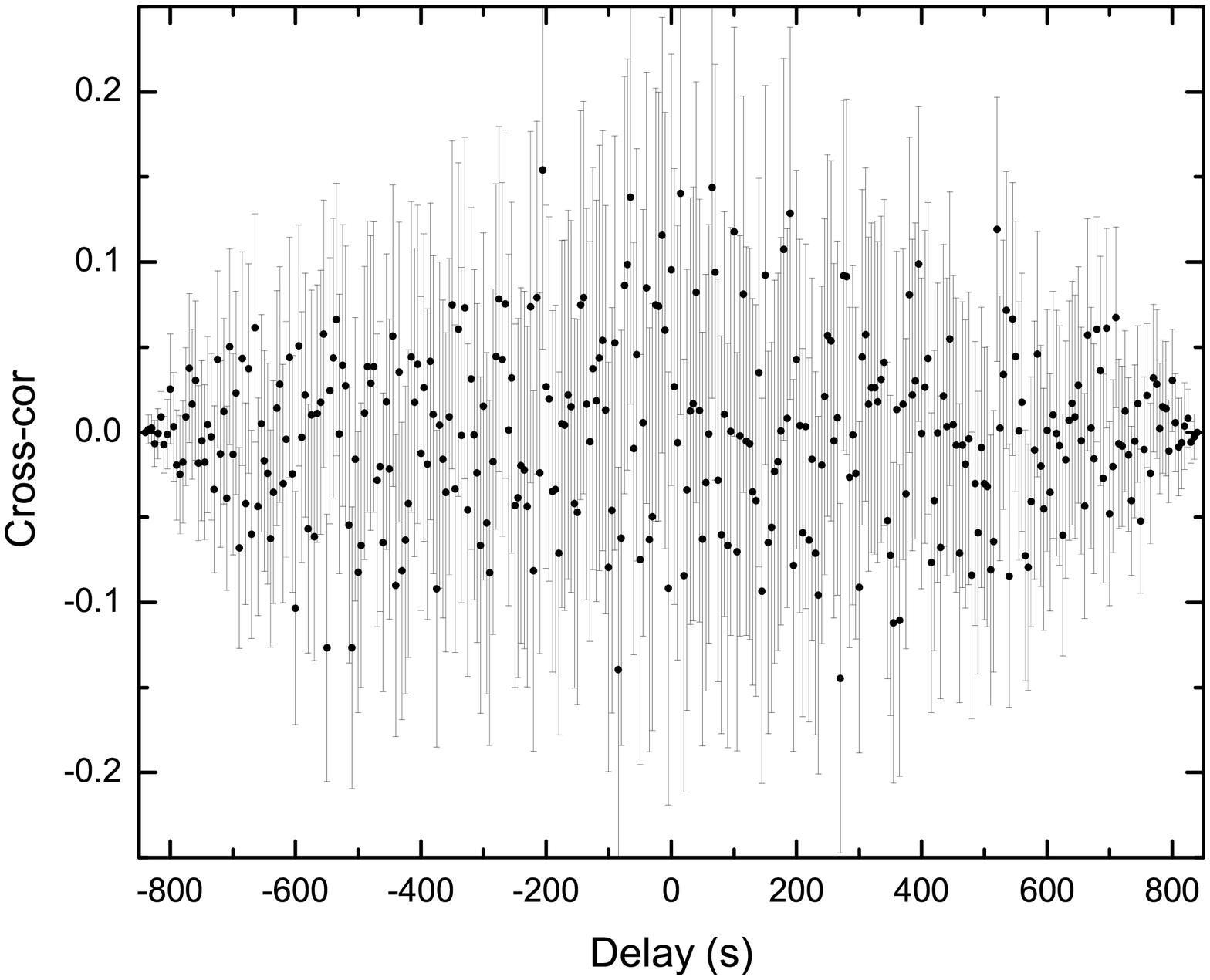}
\caption{Same as Figure 5 but for the event MJD 54352.\label{fig15}}
\end{figure}

\begin{figure}
\includegraphics[width=84mm]{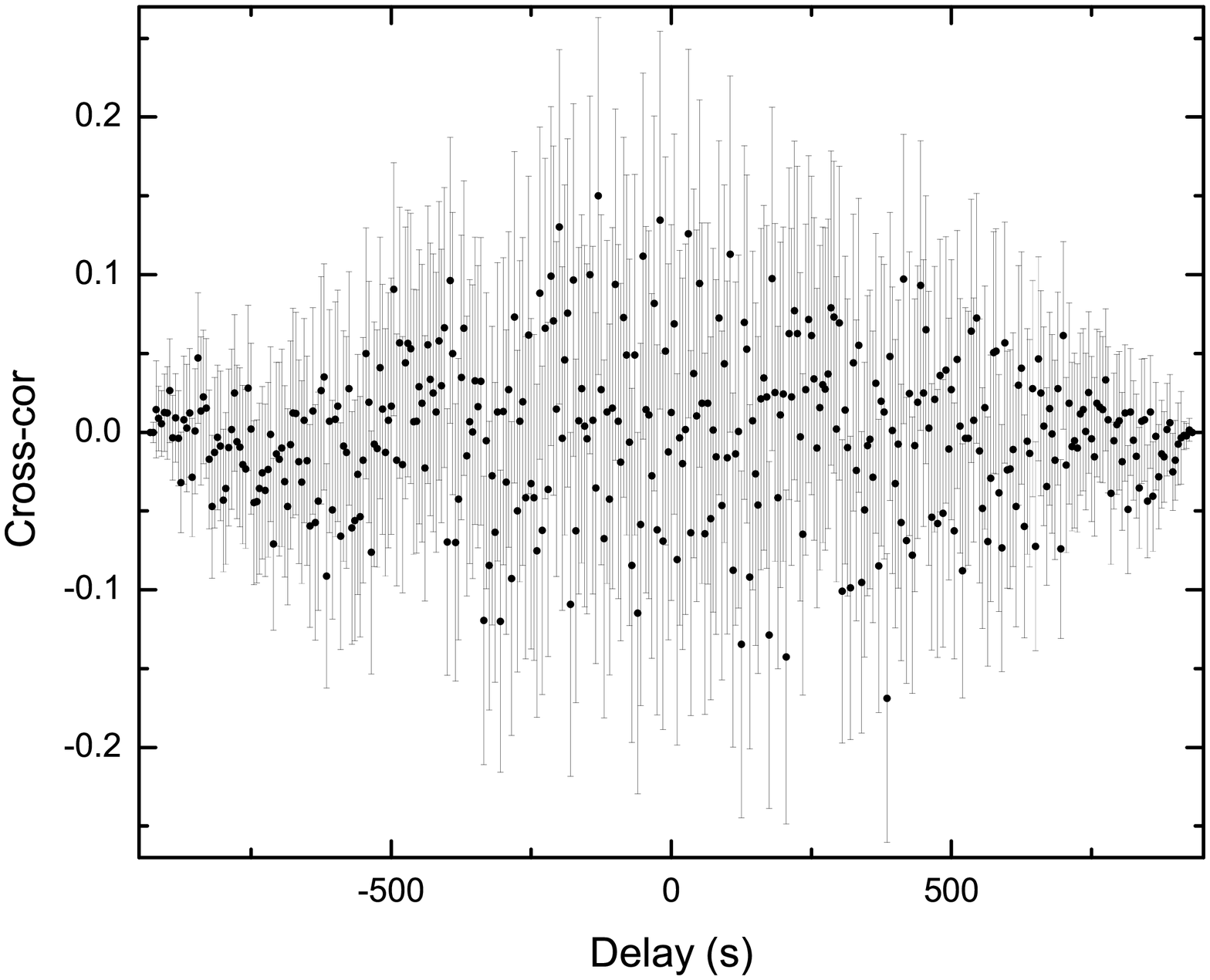}
\caption{Same as Figure 5 but for the event MJD 54376.\label{fig16}}
\end{figure}

\begin{figure}
\includegraphics[width=84mm]{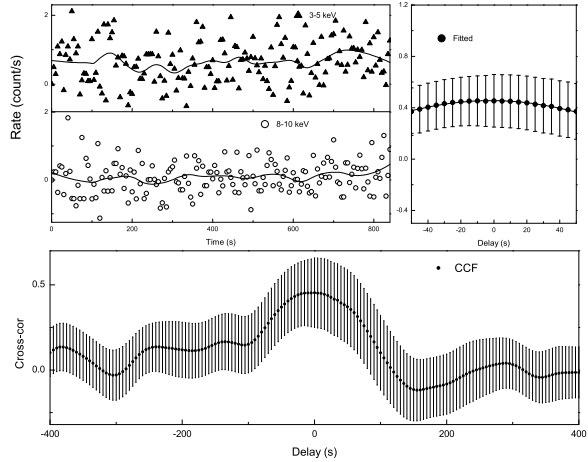}
\caption{Light curves of the event MJD 54352 at two energy bands, 3-5 keV and 8-10 keV with Loess smoothing (the solid line). The CCF after Loess smoothing is shown at the below whereas the Gaussian fitting of the peak of the CCF is shown at right side .\label{fig17}}
\end{figure}

\section{Discussion}

A model of the $LSI +61^{o} $ 303 which reproduces the energy spectrum of the system also should be able to give the spectral lags behavior obtained here from the RXTE observations. The spectral data of the source is well described by a simple featureless absorbed power law, including the flaring events. Note that such spectral behavior (power law spectrum and negative lag) has been observed for galactic (candidate) black hole binaries (in hard states) and also for AGNs \cite{utt10}.

As mentioned before there are mainly two competing models for LSI $LSI +61^{o} $ 303, the accretion driven (micro-quasar) and binary pulsar model. In the former (accretion driven) scenario, fed by the (companion) stellar wind-fed accretion disk a jet is formed in which electrons are accelerated. These electrons are cooled down by the synchrotron process or by inverse Compton scattering of local stellar photons. In the pulsar model the shock between the pulsar and the massive companion star winds is responsible for acceleration of electrons. In the microquasar model of $LSI +61^{o} $ 303 the observed \lq miniflares,\rq are due to fast variations in the stellar wind \cite{smi09, bos05}. A higher electron acceleration efficiency in the jets (and higher flux) is expected if there is a sudden increases in accretion. This leads to a higher peak of the electron energy distribution and hence to a harder emission spectra. To explain flares and flux variability on ks time-scales in the framework of binary pulsar model of the system, the clumsiness of the stellar wind is usually assumed; when the pulsar comes very close to the companion Be star and thereby is heavily exposed to the fast polar wind, clumps of relativistic electrons in the polar wind get mixed with the pulsar wind and are accelerated at the pulsar wind termination shocks \cite{zdz10, smi09}. 

The observed spectral lag features for flaring events suggest that inverse Compton scattering might be operative in the system which accommdates both the stated models. Since the time of rise and fall of the flares is just few tens of seconds and doubling time is as small as 2 seconds \cite{smi09}, the observed variability of the system demands the emission region (D) should be less than $6 \times 10^{10}$ cm unless originated relativistically. An important point is whether within this time-scale, clumps can cool electrons via inverse Compton process. In the binary pulsar model the escape time-scale of the pulsar wind mixed with the stellar wind is $t_{esc} \sim D/v_{wind}$ where $v_{wind}$ can be taken equal to the terminal velocity of the equatorial wind which is typically few $100 \; km s^{-1}$ \cite{rea10}. A detailed analysis suggest that the inverse Compton losses of about 10 MeV electrons, which will lead to inverse Compton emission of X-ray photons between 1 and 10 keV, is likely to dominate at the inner region of the system unless the magnetic field in the inner region is not too high from its Pulsar wind nebula value of $\sim 1$ G \cite{rea10}. In the outer part synchrotron emission is expected to lead and contribute dominantly to the overall X-ray emission from the system \cite{rea10}. The observed negative spectral lag of flares in the system support such a picture if the flares are generated in the inner part of the system. In the framework of microquasar model the observed flaring radiation could be due to emission from hot spots sitting above the black hole (compact object) while steady state emissions are due to the jets.  

One also cannot rule out the possibility that the flaring events observed by the RXTE mission are not associated with the $LSI +61^{o} $ 303 system \cite{li11}. Because of large field of view of PCA detector of LAXPC such a probability is there. The observed significantly different spectral lags during flaring episodes is consistent with such a scenario of different source of origin of flaring and steady emission states from the direction of the system.   

\section{Conclusion}
A study of spectral lag provides important understanding of the radiation mechanisms operating in an astrophysical object (for instance \cite{roy14}). The spectral lag study of $LSI +61^{o} $ 303 provides few very interesting results. From the RXTE observed data, we found for the first time that during the flares, low energy variations lead the higher energy variations i.e. spectral lag is negative whereas there is almost no time lag during the steady (non-flaring) states though accuracy of spectral time lags in steady state are quite limited. For flares of hundred seconds or so, spectral lags of few tens of seconds between emission of $3-5$ keV and $8-10$ keV should be considered as quite large. Any viable model of $LSI +61^{o} $ 303 needs to explain such spectral lags behavior of the system.       

In a recent work it has been found that periodic radio flares from the system lag the x-ray flare by a phase of nearly 0.2 which is equivalent to a delay of $4.58 \times 10^{5}$ s. Such a feature is explained in terms of the time of flight of energetic electrons from the binary system to radio emission region \cite{cher12}. But the spectral lag found in this work between emission of $3-5$ keV and $8-10$ keV cannot be ascribed to time of flight delay but is most likely concerned with the emission mechanism and thus the information should be helpful in understanding the nature of the source. 

%\normalem 
{\bf Acknowledgement}%\acknowledgments
The authors are thankful to an anonymous reviewer for insightful comments and suggestions that help us to improve the manuscript.  
%\end{acknowledgements}

\label{lastpage}
\end{document}